\documentclass[12pt]{iopart}
\usepackage[latin1]{inputenc}
\usepackage[dvips]{graphicx}
\usepackage[dvips]{color}
\usepackage{cite}

\usepackage{bbm}

\usepackage{iopams}
\usepackage{subfigure}
\usepackage{wasysym}
\usepackage{bbold}

\newcommand{\eqref}[1]{(\ref{#1})}

\begin{document}
\title[Kicked Bose-Hubbard systems and kicked tops]{Kicked Bose-Hubbard systems and kicked tops -- destruction and 
stimulation of tunneling}
\author{M P Strzys, E M Graefe and H J Korsch}
\address{FB Physik, Technische Universit\"at Kaiserslautern, D--67653 Kaiserslautern, Germany}
\ead{korsch@physik.uni-kl.de}

\begin{abstract}
In a two-mode approximation, Bose-Einstein condensates (BEC) in a double-well potential can be described by a many 
particle Hamiltonian of Bose-Hubbard type. We focus on such a BEC whose interatomic interaction strength is modulated 
periodically by $\delta$-kicks which represents a realization of a kicked top. In the (classical) mean-field 
approximation it provides a rich mixed phase space dynamics with regular and chaotic regions. By increasing the 
kick-strength a bifurcation leads to the appearance of self-trapping states localized on regular islands. This 
self-trapping is also found for the many particle system, however in general suppressed by coherent many particle 
tunneling oscillations. The tunneling time can be calculated from the quasi-energy splitting of the 
corresponding Floquet states. By varying the kick-strength these quasi-energy levels undergo both avoided and even 
actual crossings. Therefore stimulation or complete destruction of tunneling can be observed for this many particle 
system.
\end{abstract}


\maketitle

\section{Introduction}

After their first experimental realization \cite{Ande95,Davi95}, Bose-Einstein condensates (BEC) have stimulated an 
enormous amount of theoretical investigations.
Especially the recent progress in confining and manipulating BECs \cite{Albi05,Bloc05} affords an opportunity to realize 
a variety of quantum mechanical models experimentally.

Since a full many particle treatment of BECs is only possible for a small number of atoms, such systems are mostly 
described in the celebrated mean-field approximation, which describes the system quite well for large particle numbers 
at low temperatures. Most previous studies focused on relatively simple models like the 
two-mode system in order to 
investigate the correspondence between many particle and mean-field description \cite{Milb97,Holt01a,Holt01b}. Most 
interestingly, the mean-field description as a large particle number limit of many particle systems is formally related 
to the usual classical limit of quantum mechanics. In a number of recent papers consequences of the classical nature of 
the mean-field approximation are discussed and semiclassical aspects are introduced 
\cite{Angl01,Mahm05,Moss06,Wu06,06semiMP}. For a two-mode system even the eigenenergies and eigenstates of the many 
particle system could be reconstructed approximately from the mean-field system in a WKB type manner with astonishing 
accuracy \cite{06semiMP}. Recent experiments with a relatively small number of particles \cite{Albi05} offer the 
opportunity to study the interplay of classical and genuine quantum behavior.

Within the theory of quantum chaos systems with a periodic time dependence play an important role. Despite their 
apparently academic character systems with a delta-type time dependence are of particular relevance. Their prominence 
is not only due to their relatively simple handling in theory, but first of all to their typical behavior. Therefore 
systems like the kicked rotor or the kicked top became standard models in this area \cite{Stoe99,Haak01}.

In the present paper we focus on an $N$-particle BEC in a double-well trap with a periodically kicked interaction 
strength. Experimentally
the particle interaction may be modulated via a Feshbach resonance, a simpler realization, however, can be achieved by 
kicking the single
particle tunneling coupling instead by varying the optical potential. As we will see later, these two systems are 
equivalent in the case of a symmetric double-well. In a two-mode approximation such a system can be described by a 
two-site Bose-Hubbard type Hamiltonian. In the case of a symmetric trap this system is in fact a realization of a 
kicked top studied, e.g., by Haake \textit{et al.}~in the context of quantum chaos \cite{Haak86, Haak01}. Considering 
this BEC context, the question if the kicked top could be realized experimentally \cite{Haak00} can be answered in a 
novel way which offers the opportunity to study aspects of the system which were not investigated in the past.

One of the most prominent features of the corresponding time independent system is the so called self-trapping effect 
\cite{Milb97}. Above a critical value of the interaction strength, the system properties change qualitatively and 
unbalanced solutions appear, favoring one of the wells. A careful discussion of this effect, the relation between 
mean-field and $N$-particle behavior as well as its control by external driving fields can be found in \cite{Holt01a}. 
In the many particle system self-trapping is suppressed by tunneling oscillations.
The same effects are also present in the kicked system. In that case the dynamics of the many particle system is much 
richer.
In fact a hole bunch of prominent features of quantum dynamical systems like coherent destruction of tunneling and 
chaos assisted tunneling can be found in this model.

In the following section we present the basic model and give a short review of the Floquet formalism. In section 
\ref{sec_MF} we investigate the mean-field dynamics and compare it to the much richer many particle dynamics in 
section \ref{sec_MP}. We conclude the paper with a detailed discussion of the tunneling behavior.

\section{Basic model and Floquet formalism}
\label{sec_model}

We focus on a two-site Bose-Hubbard system with the Hamiltonian
\begin{equation} \label{ham1}
  H = \frac{\varepsilon}{2}\left(a_1^{\dagger}a_1 - a_2^{\dagger}a_2\right) + 
  \frac{v}{2}\left(a_1^{\dagger}a_2 + a_2^{\dagger}a_1\right) - \frac{c}{4} 
  \left( a_1^{\dagger}a_1 - a_2^{\dagger}a_2\right)^2,
\end{equation}
where $a_j$, $a_j^\dagger$ are bosonic
particle annihilation and creation operators for the $j$th mode,
$\varepsilon$ is the on-site energy difference, $v$ controls the single particle tunneling and $c$
the interaction strength. The Hamiltonian commutes with the number operator
\begin{equation} \label{Num}
N=a_1^{\dagger}a_1 +a_2^{\dagger}a_2
\end{equation}
so that the total number of particles $N$ is conserved.
It is convenient to introduce angular momentum operators according to the Schwinger representation
\begin{eqnarray}\label{L}
  L_x &=& \frac{1}{2}\left(a_1^{\dagger}a_2 + a_2^{\dagger}a_1\right) \nonumber \\
  L_y &=& \frac{1}{2{\rm i}}\left(a_1^{\dagger}a_2 - a_2^{\dagger}a_1\right) \\
  L_z &=& \frac{1}{2}\left(a_1^{\dagger}a_1 - a_2^{\dagger}a_2\right),\nonumber
\end{eqnarray}
which obey the $su(2)$ commutation relations $[L_i,L_j]={\rm i}L_k$, where
$i,j,k = x,y,z$ and cyclic permutations.
The angular momentum quantum number is equal to $\ell=N/2$. In terms of these operators the Hamiltonian \eqref{ham1} 
assumes the form
\begin{equation}\label{Hx}
  H = \varepsilon L_z + vL_x - cL_z^2.
\end{equation}

As already discussed in the introduction, we now consider a periodically kicked system, in particular the case of
a kicked interaction strength
\begin{equation}\label{ct}
c(t) = c\,\tau\sum_{m=-\infty}^{+\infty} \delta(t-m\tau), 
\end{equation}
i.e.~the time-periodic Hamiltonian
\begin{equation}\label{H}
  H(t) =  H_0 + c(t)V 
 \ ,\quad H_0=\varepsilon L_z + vL_x,\ ,\quad  V=-L_z^2.
\end{equation}
In the symmetric case $\varepsilon = 0$ this is in fact the Hamiltonian of a
kicked top \cite{Haak86, Haak01}. A similar system with kicked single particle
tunneling coupling $v$ has been studied in \cite{Xie05}. Mathematically, the 
benefit of a kicked system is the enormous simplification of the dynamics.
The mean-field approximation, for example, can be evaluated in closed form as 
outlined in section \ref{sec_MF}.

Since the Hamiltonian \eqref{ham1} is periodic in time, $H(t+\tau) = H(t)$, it is useful to introduce the propagator over 
one period, the so called Floquet-operator
\begin{equation}\label{floq}
F(t) = U(t+\tau,t) = \mathcal T \left[\exp{\left(-{\rm i}\int_t^{t+\tau} H(t'){\rm d}t'\right)}\right]
\end{equation}
where the operator $\mathcal T$ takes care of the right time order. The Floquet-operator \eqref{floq} is by definition 
unitary and thus has unimodular eigenvalues
\begin{equation}\label{k3_eigenwertgl}
F(t)\left|\kappa(t)\right> = {\rm e}^{-{\rm i}\epsilon_{\kappa}\tau}\left|\kappa(t)\right>.
\end{equation}
The $N+1$ quasi-energies $\epsilon_\kappa$ are the natural counterparts of the energies of time-independent systems. 
We choose $\epsilon_\kappa$ in the first Brillouin-zone $[-\pi/\tau,\pi/\tau)$.

The eigenstates of the Floquet-operator, the so called Floquet-states $\left|\kappa(t)\right>$, form an orthonormal 
basis of the Hilbert space and obey
\begin{equation}\label{k3_Feigenzu}
\left|\kappa(t)\right> = {\rm e}^{-{\rm i}\epsilon_\kappa t}\left|\phi_\kappa(t)\right>\quad {\rm with}\quad 
\left|\phi_\kappa(t+\tau)\right> = \left|\phi_\kappa(t)\right>.
\end{equation}

For the $\delta$-type time dependence the Floquet-operator factorizes
\begin{equation}\label{floq2}
F = F(0) = {\rm e}^{{\rm i}cL_z^2\tau}{\rm e}^{-{\rm i}(\varepsilon L_z + vL_x)\tau}.
\end{equation}
Similar expressions would be obtained by modulating the single particle tunneling coupling $v$ and the on-site energy 
difference $\varepsilon$ with periodic $\delta$-kicks instead of the particle interaction $c$. In this case one would 
arrive at a Floquet-operator
\begin{equation}
\tilde F = {\rm e}^{-{\rm i}(\varepsilon L_z + vL_x)\tau}{\rm e}^{{\rm i}cL_z^2\tau},
\end{equation}
where the order of the two exponential factors of \eqref{floq2} is switched. The stroboscopic dynamics induced by 
iterations of the Floquet-operator, however, is not changed.

The additional symmetry (in the symmetric case for $\varepsilon = 0$) can be expressed via an invariance 
under rotations $R_x=R_x(\pi)={\rm e}^{-{\rm i}\pi L_x}$ around the $x$-axis about $\pi$. The Floquet-operator commutes with $R_x$,
\begin{equation} \label{k3_FRx}
[F,R_x] = 0,
\end{equation}
and the Floquet states can be separated into two symmetry classes, one even and one odd under the rotation $R_x$. If 
$N$ is even, the angular momentum quantum number $\ell$ is an integer. Thus we have $R_x^2 = \mathbbm 1$ and the eigenvalues 
of $R_x$ are $\pm 1$. Equation \eqref{k3_eigenwertgl} can be reformulated according to
\begin{equation} \label{k3_Feig}
F\left|\kappa_\pm\right> = {\rm e}^{-{\rm i}\epsilon_{\kappa}\tau}\left|\kappa_\pm\right>
\end{equation}
where
\begin{equation}
R_x\left|\kappa_\pm\right> = \pm \left|\kappa_\pm\right>.
\end{equation}
If $N$ is odd $R_x^2 = -\mathbbm 1$ holds, since $\ell$ is half an integer. The eigenvalues of $R_x$ 
then are $\pm{\rm i}$. 
This yields as well two classes of Floquet states $\left|\kappa\right>$, which obey
\begin{equation}
R_x\left|\kappa_\pm\right> = \pm {\rm i} \left|\kappa_\pm\right>.
\end{equation}
These two symmetry classes will be of importance for the tunneling process investigated in the following.

\section{Mean-field dynamics}
\label{sec_MF}

The celebrated mean-field approximation can be achieved by replacing the field operators by complex numbers 
\begin{equation} \label{k1_ersetzung}
a_j \longrightarrow \psi_j, \quad a_j^{\dagger} \longrightarrow \psi_j^*, \quad j = 1,2.
\end{equation}
Note, however, that the mapping \eqref{k1_ersetzung} is not well-defined, 
since complex numbers commute in contrast to operators. 
Therefore, the operators should be rewritten in symmetrized form in $a_j$ and $a_j^{\dagger}$ before
the replacement \eqref{k1_ersetzung} is carried out \cite{Moss06,06semiMP}.
Here the Hamiltonian \eqref{ham1} is already symmetric, however the number operator \eqref{Num} appears
as
\begin{equation} \fl
N=a_1^{\dagger}a_1 +a_2^{\dagger}a_2=\frac{a_1^{\dagger}a_1 + a_1a_1^{\dagger}+ a_2^{\dagger}a_2 + a_2a_2^{\dagger}-2}{2}
\ \longrightarrow \ \left|\psi_1\right|^2+\left|\psi_2\right|^2-1
\end{equation}
and  the mean-field wave function is normalized according to $|\psi_1|^2 + |\psi_2|^2 = N + 1$ (see also \cite{06semiMP}).

This is in fact a classical limit of the system, equivalent to the limit $1/N \rightarrow 0$, where the parameter $1/N$ 
takes over the role of an effective Planck constant \cite{Haak86,Zhan90}. 
In the case of the Hamiltonian \eqref{ham1} this leads to a Hamiltonian function
\begin{equation}\label{Hmean}\fl
\mathcal{H}(\psi_{1,2},\psi_{1,2}^*) = \frac{\varepsilon}{2}(\psi_1^*\psi_1-\psi_2^*\psi_2)+\frac{v}{2}(\psi_1^*\psi_2+\psi_2^*\psi_1)-\frac{c}{4}(\psi_1^*\psi_1-\psi_2^*\psi_2)^2
\end{equation}
of the conjugate variables $\psi_j$ and $\psi_j^*$ which obey the classical canonical equations of motion
\begin{equation}
{\rm i}\frac{{\rm d}\psi_{j}}{{\rm d}t}=\frac{\partial\mathcal{H}}{\partial\psi_j^*}, \quad j = 1,2.
\end{equation}

The dynamics of such a system can be equivalently put in the form of a discrete Gross-Pitaevskii equation (GPE), or 
discrete nonlinear Schr\"odinger equation, in terms of the nonlinear matrix equation
\begin{equation}\label{GPE}
{\rm i}\frac{{\rm d}}{{\rm d}t}\left( \begin{array}{c}\psi_{1}\\\psi_{2}\end{array}\right)=\frac{1}{2}\left( 
\begin{array}{cc}
\varepsilon+c\kappa &  v\\
v & -(\varepsilon+c\kappa)
\end{array}\right)\left( \begin{array}{c}\psi_{1}\\\psi_{2}\end{array}\right),
\end{equation}
where $\kappa=|\psi_{2}|^{2}-|\psi_{1}|^{2}$ is the population imbalance of the two wells. The dynamics described by 
this equation of
motion conserves the norm of the wave function and can be conveniently visualized by a mapping onto the Bloch sphere: 
Similar to
\eqref{L}, one can define the three quantities
\begin{equation}\label{blochvek}\fl
s_x=\frac{1}{2}(\psi_1^*\psi_2+\psi_1\psi_2^*), \quad s_y=\frac{1}{2{\rm i}}(\psi_1^*\psi_2-\psi_1\psi_2^*), 
\quad s_z=\frac{1}{2}(|\psi_1|^2-|\psi_2|^2),
\end{equation}
the components of the Bloch vector $\boldsymbol s \in \mathbbm R^3$. The norm of the Bloch vector 
$s = |\boldsymbol s| = (N+1)/2=\ell +1/2$ 
is determined by the normalization of the mean-field wave function. Writing the GPE \eqref{GPE} in 
terms of the Bloch vector yields the nonlinear Bloch equations
\begin{eqnarray}\label{nlbloch}
\dot s_x & = & -\varepsilon s_y + 2cs_y s_z\nonumber\\
\dot s_y & = & \varepsilon s_x - v s_z -2cs_x s_z\\
\dot s_z & = & v s_y \nonumber.
\end{eqnarray}
For the periodically kicked system \eqref{H} one can, 
as in the many particle case, define a Floquet operator $\mathcal{F}$ for the mean-field Bloch vector.
Here again the Floquet operator factorizes: For $c(t)=0$ the dynamics is linear and can be solved in closed form.
The resulting time evolution can be described by a matrix
\begin{equation}\label{k2_R(t)}
\mathcal{R} = \left( \begin{array}{ccc}
\frac{v^2}{\omega^2}+\frac{\varepsilon^2}{\omega^2}\cos{\omega \tau} & -\frac{\varepsilon}{\omega}\sin{\omega \tau} &  
\frac{\varepsilon v}{\omega^2}\left( 1-\cos{\omega \tau}\right) \\
\frac{\varepsilon}{\omega}\sin{\omega \tau} & \cos{\omega \tau} & -\frac{v}{\omega}\sin{\omega \tau} \\
 \frac{\varepsilon v}{\omega^2}\left( 1-\cos{\omega \tau}\right)  & \frac{v}{\omega}\sin{\omega \tau} & 
 \frac{\varepsilon^2}{\omega^2} + \frac{v^2}{\omega^2}\cos{\omega \tau}
\end{array}\right),
\end{equation}
a rotation about the axis $(v/\omega,0,\varepsilon/\omega)$ with frequency
\begin{equation}\label{omega}
\omega =\sqrt{\varepsilon^2+v^2}.
\end{equation}
This rotation is followed by a kick, which conserves the $z$-component $s_z$, described by the torsion matrix
\begin{equation}
\mathcal{K} = \left( \begin{array}{ccc}
\cos{2c\tau s_z } & \sin{2c\tau s_z } & 0 \\
-\sin{2c\tau s_z } & \cos{2c\tau s_z } & 0 \\
0 & 0 & 1 \end{array}\right).
\end{equation}
The full mean-field dynamics therefore reads
\begin{equation}\label{Fs}
\boldsymbol s_{m+1} = \mathcal{F}\boldsymbol s_{m} = \mathcal{K}\,\mathcal{R}\,\boldsymbol s_{m},
\end{equation}
where $\boldsymbol s_m = \boldsymbol s(m\tau)$, $m \in \mathbbm Z$.

\begin{figure}
\begin{center}
   \includegraphics[scale=1]{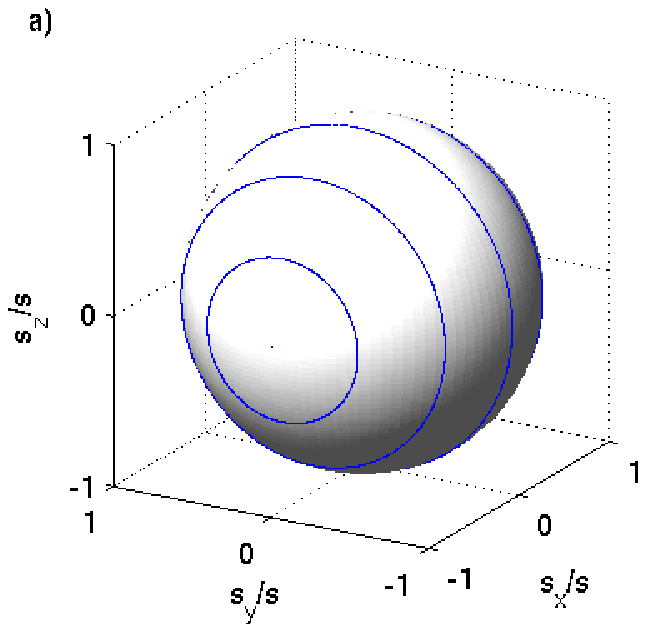}
   \includegraphics[scale=1]{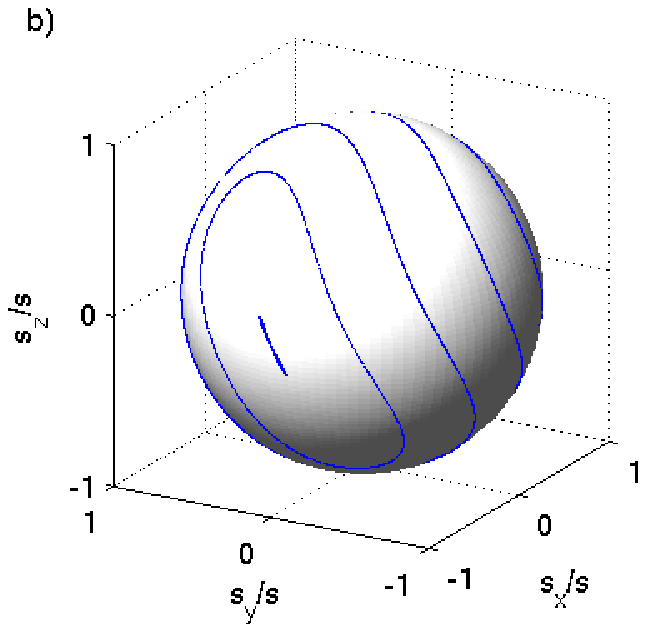}
   \includegraphics[scale=1]{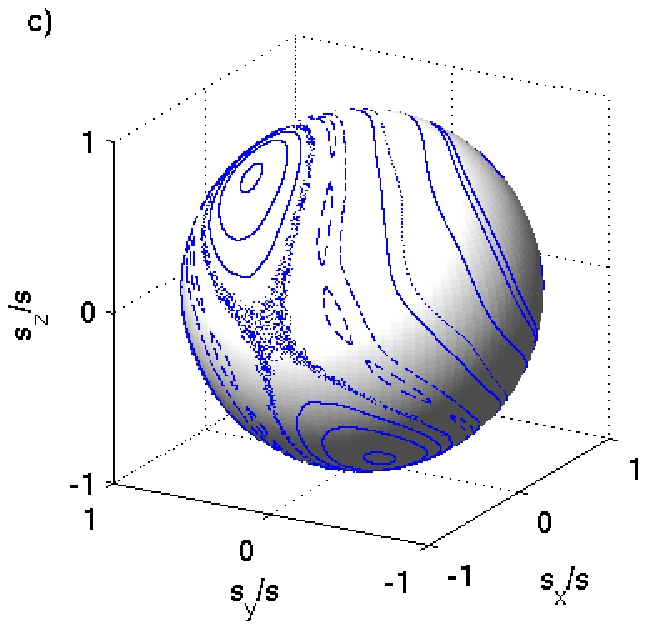}
   \includegraphics[scale=1]{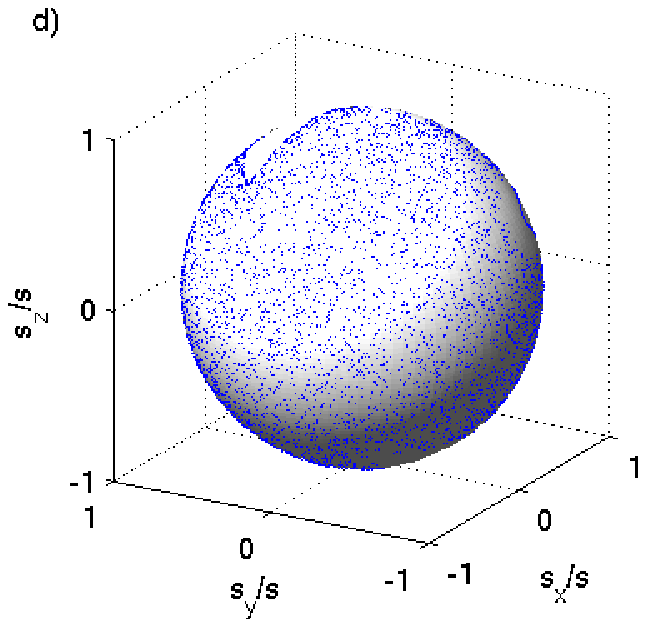}
   \caption{Stroboscopic dynamics of the Bloch vector $\boldsymbol{s}$ for the parameters  $\varepsilon=0$, $v=1$, 
   $\tau=1$ and \mbox{a) $c = 0$},  b) $c=1.1/(N+1)$,  c) $c=1.8/(N+1)$ and d) $c=3.5/(N+1)$, respectively.}
   \label{Figure:kug_1}
\end{center}
\end{figure}

In the following numerical studies we will consider a symmetric system ($\varepsilon = 0$).
Stroboscopic iterations of $\boldsymbol s$ according to the discrete mapping \eqref{Fs} 
with $v = 1$ and four different values of the interaction-strength $c$ are presented in figure \ref{Figure:kug_1}.
Depending on the inter particle interaction strength $c$, the system exhibits the typical behavior of classical 
nonlinear and thus chaotic systems. In the case of no interaction ($c = 0$) shown in figure \ref{Figure:kug_1} a) 
the system is integrable and one observes the usual Rabi oscillations which give rise to simple rotations of the 
Bloch vector. At the two stable fixed points $(\pm s, 0, 0)$ the population imbalance of the two wells is constant 
in time. With increasing interaction the linear rotation around the $x$-axis is more and more perturbed by a torsion 
along the $z$-axis and bifurcations of the fixed points occur. For $c=1.1/(N+1)$, shown in figure \ref{Figure:kug_1} 
b), the first bifurcation of the fixed point $(-s,0,0)$ has just caused two new fixed points to appear. Increasing $c$ 
further, the regular islands around the new fixed points grow while the fixed points wander toward the poles of the 
Bloch sphere; the example $c = 1.8/(N+1)$ can be seen in figure \ref{Figure:kug_1} c). In addition, chaotic regions 
appear in the vicinity of the unstable fixed point. The two regular islands will be of special interest in the following, 
since they correspond to the famous self-trapping states of the non-kicked system as they are localized on the northern 
and southern hemisphere, respectively. Unlike the non-kicked system, self-trapping occurs only in a certain regime,
since increasing $c$ even further leads inevitably to global chaos. The case $c = 3.5/(N+1)$ is shown in figure
\ref{Figure:kug_1} d); the chaotic sea spreads noticeably over almost the whole sphere, sparing only small regular
islands.

The critical value of the parameter $c$ for the occurrence of self-trapping, i.e.~the point of bifurcation of the
prominent fixed point $(-s, 0, 0)$, can be calculated in dependence on the other system parameters by linearization.
To guarantee stability of the fixed point $(-s,0,0)$ the interaction strength must obey the inequalities
\begin{equation}
\frac{\mp 1 - \cos{v\tau}}{s \sin{v\tau}} <  c  < \frac{\pm 1-\cos{v\tau}}{s \sin{v\tau}}\quad \textrm{for}\quad
\sin{v\tau}\gtrless 0.
\end{equation}
Thus, for parameters $\tau = v = 1$, this fixed point is stable for $c\in(-3.661,1.092)/(N+1)$.

\section{Many particle quantum dynamics}
\label{sec_MP}

In order to compare the mean-field dynamics with the full quantum dynamics, we introduce the $SU(2)$ coherent states,
also called atomic
coherent states $\left|\vartheta,\varphi\right>$. They are constructed by  an arbitrary $SU(2)$-rotation
$R(\vartheta,\varphi) =
\exp{\left({\rm i}\vartheta(L_x \sin{\varphi} - L_y\cos{\varphi})\right)}$ of the extremal Fock state $\left|N\right>$
with all particles in the first well, which is a coherent state located at the north pole of the Bloch sphere
\cite{Arec72, Gilm75, Pere86, Zhan90},
\begin{equation}
\left|\vartheta,\varphi\right> = R(\vartheta,\varphi)\left|N\right>.
\end{equation}
In the Fock basis they take the form
\begin{equation}
\left|\vartheta,\varphi\right> = \sum_{n=0}^{N}\sqrt{\left({N}\atop{n}\right)}\cos^n{\left(\frac{\vartheta}{2}\right)}
\sin^{N-n}{\left(\frac{\vartheta}{2}\right)}{\rm e}^{{\rm i}(N-n)\varphi}\left|n\right>,
\end{equation}
where
\begin{equation}
  \left|n\right>=\left|n,N-n\right>=\frac{1}{\sqrt{n!(N-n)!}}{(a_1^{\dagger})}^n {(a_2^{\dagger})}^{N-n}\left|0,0\right>,
\end{equation}
with $n = 0, 1, \dots , N$, are the usual Fock states. The expectation values of the angular momentum operators in a
coherent state $\left<L_j\right> = \left<\vartheta,\varphi\right|L_j\left|\vartheta,\varphi\right>$, $j = x,y,z$ are
given by
\begin{equation}
\left<L_x\right> = \frac{N}{2}\sin{\vartheta}\cos{\varphi},\quad \left<L_y\right> = \frac{N}{2}\sin{\vartheta}
\sin{\varphi},\quad \left<L_z\right> = \frac{N}{2}\cos{\vartheta}.
\end{equation}
Thus, the vector $\left<\boldsymbol L\right> = \left( \left<L_x\right>,\left<L_y\right>,\left<L_z\right> \right)$
points on a sphere with
radius $\ell = N/2$.

Here again, the Floquet operator induces a discrete mapping from kick to kick. In the Heisenberg picture this yields
the mapping
\begin{equation}\label{k3_Nteilchenrekursion}
L_j^{(m+1)} = F^\dagger L_j^{(m)}F = {\rm e}^{{\rm i}H_0\tau}\left({\rm e}^{{\rm i}cV\tau}L_j^{(m)}{\rm e}^{-{\rm i}cV\tau}
\right){\rm e}^{-{\rm i}H_0\tau}
\end{equation}
for the angular momentum operators $L_j^{(m)}$ after the $m$-th kick.
In the symmetric case  with $\varepsilon=0$ these equations assume the simple form
\begin{eqnarray}
F^\dagger L_xF &=& \frac{1}{2}\left[\left( L_x+{\rm i}\left(L_y\cos{v\tau}-L_z\sin{v\tau}\right) \right)
{\rm e}^{-{\rm i}c(1+2L_y\sin{v\tau}+2L_z\cos{v\tau})}\right] \nonumber \\ & & \quad + \quad\textrm{h.c.} \nonumber\\
F^\dagger L_yF &=& \frac{1}{2{\rm i}}\left[\left(L_x+{\rm i}\left(L_y\cos{v\tau}-L_z\sin{v\tau}\right)\right)
{\rm e}^{-{\rm i}c(1+2L_y\sin{v\tau}+2L_z\cos{v\tau})}\right] \nonumber \\ & & \quad + \quad\textrm{h.c.} \nonumber \\
F^\dagger L_zF &=& L_y\sin{v\tau} + L_z\cos{v\tau}.
\end{eqnarray}

Now we can compare the many particle with the mean-field dynamics. For this purpose we focus on the normalized Bloch
vector $\boldsymbol
s/s$ and the normalized vector of expectation values $\left<\boldsymbol L\right>/\ell$ of the angular momentum operators.
In the classically regular regime, the full quantum dynamics closely resembles the mean-field system and shows the usual
Rabi oscillations according to the single particle tunneling, which are modulated by a  breakdown and revival scenario
due to the discrete spectrum of the system. A similar behavior is very common
and known for many systems \cite{Milb97,Holt01b}.
\begin{figure}
	\centering
	\includegraphics[scale=1]{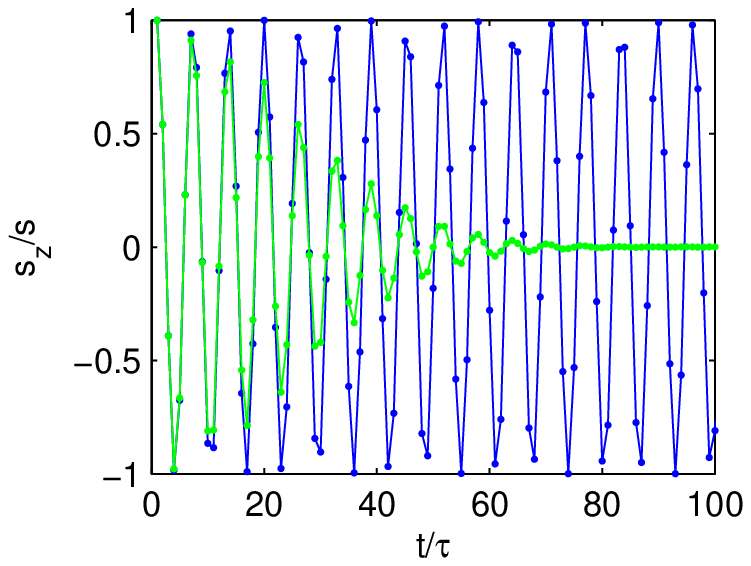}\includegraphics[scale=1]{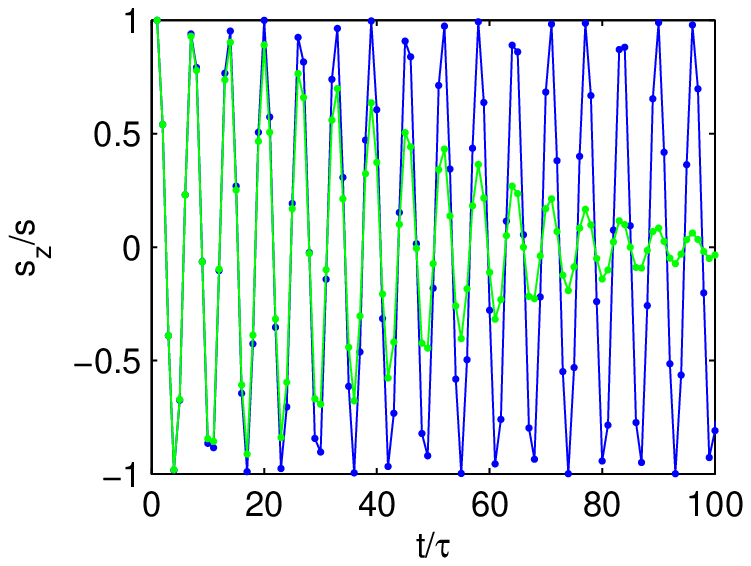}
	\caption{Time evolution of the $z$-component of the angular momentum expectation value for $\varepsilon = 0$, $v =1$ and
	$c=0.5/(N+1)$. The many particle dynamics ($\textcolor{green}{\bullet}$ ) for an initial  coherent state
	$\left|N\right>$ located at the north pole is compared with the mean-field trajectory  ($\textcolor{blue}{\bullet}$)
	for a starting vector $\boldsymbol s/s = \boldsymbol x = (0,0,1)$  (the lines are plotted to guide the eye); left: $N = 50$
	particles, right: $N = 100$ particles.}
	\label{Figure:mean-Nteil_c_0-5_N_33}
\end{figure}
Figure \ref{Figure:mean-Nteil_c_0-5_N_33} shows a comparison of the mean-field and many particle
dynamics. Shown is the $z$-component of the angular momentum expectation value (i.e.~the population imbalance of the two wells)
 for a symmetric system ($\varepsilon = 0$)
with $v=1$ and a weak interaction strength $c = 0.5/(N+1)$, where the initial state is a coherent state
$\left|N\right>$ located at the north pole, or $\boldsymbol s/s = (0,0,1)$ in the mean-field approximation.
Both oscillate with frequency $\omega=1$ (compare \eqref{omega}) and the envelope of the
many particle expectation values decay with a width $\sim \sqrt{N+1}$.
Note, however, that the deviation between the mean-field and many particle dynamics is due to the fact
that the mean-field description is based on a single trajectory and can be cured to some extent by propagating an
ensemble approximating initially the coherent state distribution of the many particle system \cite{07meanf}.

\begin{figure}
	\centering
	\includegraphics[scale=1]{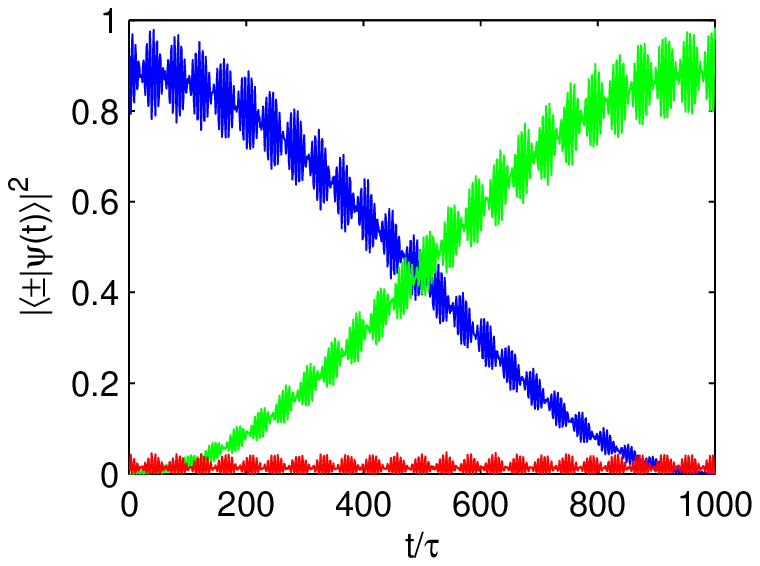}
	\caption{Squared absolute values of the projections of $\left|\psi(t)\right>$ on the coherent states
	$\left|\pm\right>$ during propagation over $1000$ periods for $N=20$ particles, $\varepsilon = 0$, $v=1$, $\tau=1$
	and $c=2/(N+1)$; $\textcolor{blue}{-}$ $\left|\left<-\big|\psi(t)\right>\right|^2$, $\textcolor{green}{-}$
	$\left|\left<+\big|\psi(t)\right>\right|^2$ and $\textcolor{red}{-}$ orthogonal subspace.}
	\label{Figure:inselproj_N_20_c_2}
\end{figure}

Our interest, however, is not focused on the regular regime, but on the mixed regular-chaotic regime with the two prominent
regular islands centered at the self-trapping states $\boldsymbol s_\pm = \boldsymbol s(\vartheta_\pm,\varphi_\pm)$. A
mean-field trajectory started in one of these states will stay there forever. The full quantum system however, shows a
much richer behavior here. Starting the propagation in one of the corresponding coherent states
$\left|\pm\right> = \left|\vartheta_\pm,\varphi_\pm\right>$ the system exhibits dynamical tunneling to the other island,
i.e.~the state $\left|\mp\right>$.
Such dynamical tunneling processes are known for many systems \cite{Grif98,Bayf99}, e.g., a particle in periodically
driven double-well potential
\cite{Lin90, Lin92, Bavl93} or a periodically driven rotor \cite{95tun,Bonc98, Aver02}. Such a system can be realized with
a BEC \cite{Hens00, Hens01a, Hens01b, Stec01, Salm02, Osov05} allowing direct experimental observation of the tunneling.
\begin{figure}
	\centering
	\includegraphics[scale=1]{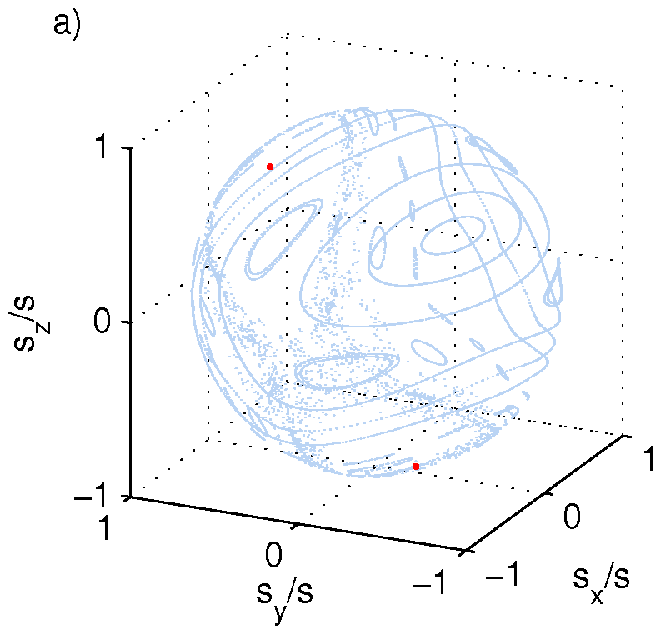}
	\includegraphics[scale=1]{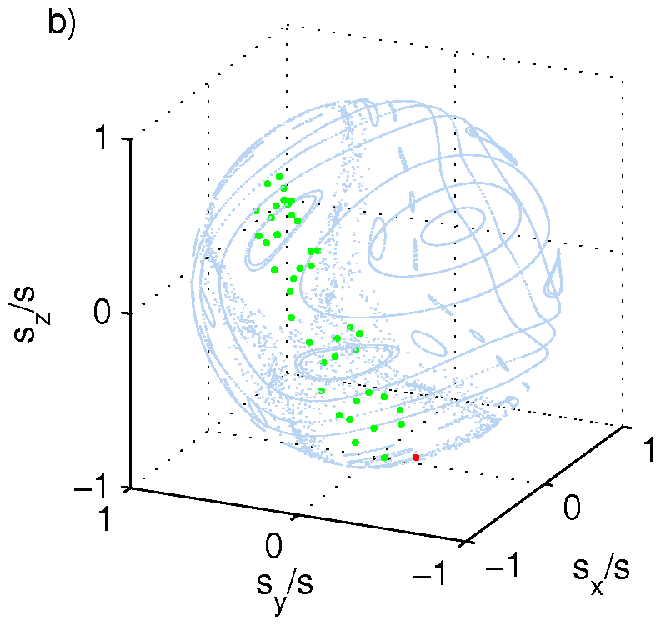}
	\includegraphics[scale=1]{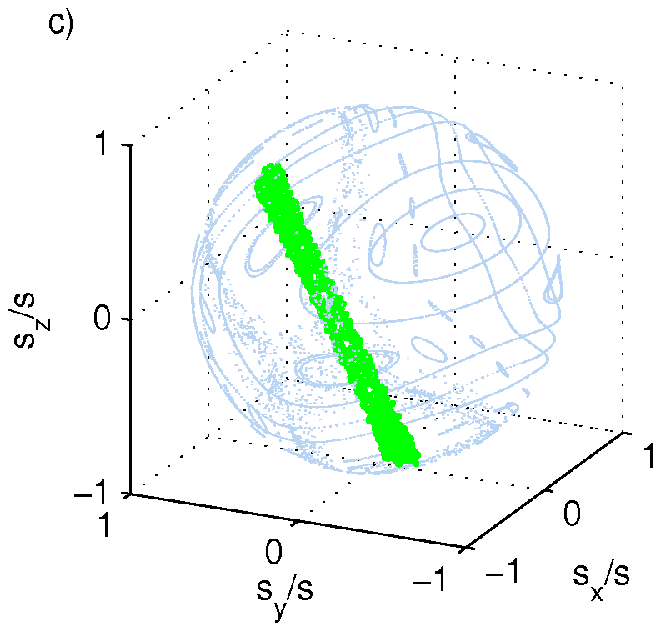}
	\includegraphics[scale=1]{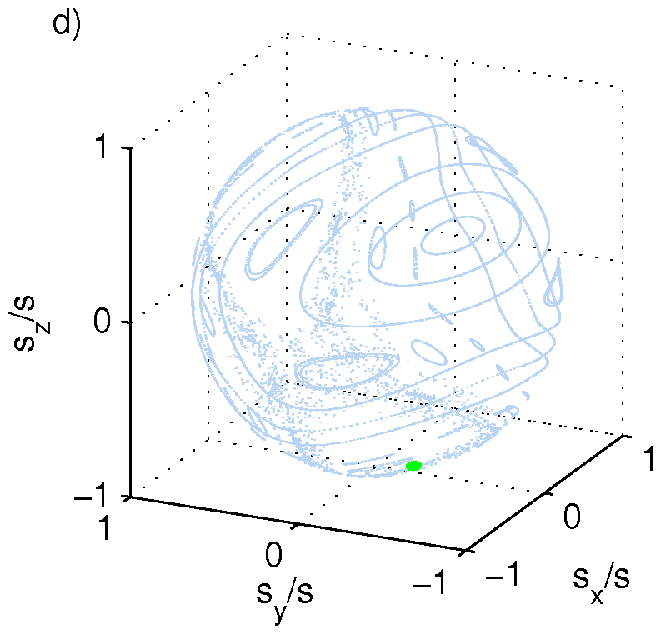}
	\caption{a) Mean-field dynamics for the parameters $\varepsilon = 0$, $v=1$, $\tau = 1$ and $c=2/(N+1)$, the
	fixed points $\boldsymbol s_\pm$ are marked with red dots. b) Propagation of many-particle state $\left|-\right>$ located on the southern fixed point for $N=10$ particles;
	the dynamical tunneling to the northern fixed point lasts $40$ periods here; c) as before, but $N = 20$ particles;
	tunneling during $1000$ periods d) as in c), but $N=100$ particles, dynamical tunneling cannot be observed here.}
	\label{Figure:tunnel_c_2_fix}
\end{figure}
It should be emphasized, that here the dynamical tunneling between both of the island states is a coherent many particle
process and may not
be confused with the single particle tunneling which causes the Rabi oscillations. Nevertheless it corresponds to an actual
tunneling of
the condensate between the potential wells, since the northern state is localized almost in the first well, whereas the
southern state
occupies mostly the second well. In figure \ref{Figure:tunnel_c_2_fix} the tunneling process starting from the fixed point
on the southern
hemisphere in the state $\left|\psi(0)\right> = \left|-\right>$ is shown for several parameter sets. An interaction strength
of $c =
2/(N+1)$ is chosen to ensure the two regular islands to be not too small and well separated. One observes dynamical
tunneling through the
Bloch sphere. The normalized expectation value $\left<\boldsymbol L\right>/\ell$ spirals its way on a tube encircling
the straight line connecting the two island states to the other side of the sphere. The radius of the tunneling tube is
determined by the uncertainty of the direction of the angular momentum vector in a coherent state, which is proportional
to $1/N$. The more particles one puts into the system, the tighter is the spiral around the connection line of the fixed
points and the longer gets the tunneling time. However, under some conditions the tunneling time can also decrease with
a growing number of particles, leading to a strongly enhanced population transfer, as will be explained in section
\ref{subsecparameter}. But let us first have a closer look at the tunneling process. The squared absolute values
$\left|\left<\psi(t)\big|\pm\right>\right|^2$ of the projections of $\left|\psi(t)\right>$ on the coherent island states
and the orthogonal subspace for $\left|\psi(0)\right>=\left|-\right>$ for case c) of figure \ref{Figure:tunnel_c_2_fix}
with $N = 20$ is shown in figure \ref{Figure:inselproj_N_20_c_2} as a function of time.
After $1000$ periods, the population has been transferred almost completely to the state $\left|+\right>$ on the
northern hemisphere.

For all particle numbers $N$ considered in figure \ref{Figure:tunnel_c_2_fix}, the population $\left|\left<\kappa\big|-\right>\right|^2$ of the southern
coherent island state on the Floquet states $\left|\kappa\right>$ is significantly large only for two of the $(N+1)$
Floquet states which are denoted by $\left|\kappa_\pm\right>$ with corresponding quasi-energies $\epsilon_\pm$. As
indicated by the $\pm$-sign, they belong to different symmetry classes and the island states can be reconstructed up
to 94\,\% by the linear combinations
\begin{equation}\label{k4_linkomb}
\left|\psi_\pm\right> = \frac{1}{\sqrt{2}}\left(\left|\kappa_+\right> \mp \left|\kappa_-\right>\right)
\end{equation}
which hence are a good approximation for the coherent states $\left|\pm\right> \approx \left|\psi_\pm\right>$.
If one chooses $\left|\psi_-\right>$ as a starting vector, $\left|\psi(t)\right>$ always stays a linear combination of
the $\left|\kappa_\pm\right>$ during propagation. Remembering equation \eqref{k3_Feigenzu} for the decomposition of the
Floquet states, the propagation over $n$ kick periods yields
\begin{eqnarray}
\left|\psi(n\tau)\right> &=& \frac{1}{\sqrt{2}}{\rm e}^{-{\rm i}n\epsilon_+\tau
}\left(\left|\kappa_+\right> +
{\rm e}^{-{\rm i}n(\epsilon_- - \epsilon_+)\tau}\left|\kappa_-\right>\right)\\
&=& \frac{1}{2}{\rm e}^{-{\rm i}n\epsilon_+\tau
}\left( \left(1 - {\rm e}^{-{\rm i}\phi_n} \right)\left|+\right> + \left(1 + {\rm e}^{-{\rm i}\phi_n} \right)\left|-\right> \right),
\end{eqnarray}
with $\phi_n = n(\epsilon_- - \epsilon_+)\tau$. The expectation value of the angular momentum varies as
\begin{equation}\label{ellipse}
\left<\boldsymbol L (n\tau) \right> = \frac{1-\cos{\phi_n}}{2}\boldsymbol L_+ + \frac{1+\cos{\phi_n}}{2}\boldsymbol L_- + \sin{\phi_n}{\rm Im}{\boldsymbol L_{-+}}
\end{equation}
with $ \boldsymbol L_\pm = \left<\pm\right|\boldsymbol L \left|\pm\right>$, $ \boldsymbol L_{-+} = \left<-\right|\boldsymbol L \left|+\right>$, where in the present cases ${\rm Im}\boldsymbol L_{-+}$ is zero or very small so that the ellipse \eqref{ellipse} degenerates to a straight line connecting the island centers at $\boldsymbol L_-$ and $\boldsymbol L_+$. Thus the spiraling on a tube around this line is due to the other contributing Floquet states.
We conclude, that the subsystem consisting of the two island states $\left|\pm\right>$ can be described in a good
approximation by a two state system of the Floquet states $\left|\kappa_\pm\right>$. Thus, as any two level system,
the system exhibits periodic tunneling with period
\begin{equation}\label{k4_tunnelzeit}
T_{\rm tunnel} = \frac{2\pi}{\epsilon_- - \epsilon_+} = \frac{2\pi}{\Delta\epsilon}.
\end{equation}
These considerations are entirely equivalent to the tunneling process through a symmetric potential barrier,
where the quasi-energies adopt the function of the energies of time independent systems. Thus the tunneling period
is solely ruled by the quasi-energy splitting $\Delta\epsilon$. One should keep in mind, that despite the similarities
in the formalism this is no single particle, but a coherent many particle process. Still the dynamical tunneling
between the two island states corresponds to an actual tunneling of the BEC from one potential well to the other.
If the system is localized in the southern island $\left|-\right>$, the expectation value of $L_z$ is negative and
the condensate is located mainly in the second well; in the case $c = 2/(N+1)$ over $80\,\%$ of the condensate is
trapped there. The opposite is true for the northern island $\left|+\right>$.

\begin{figure}
	\centering
	\includegraphics[scale=1]{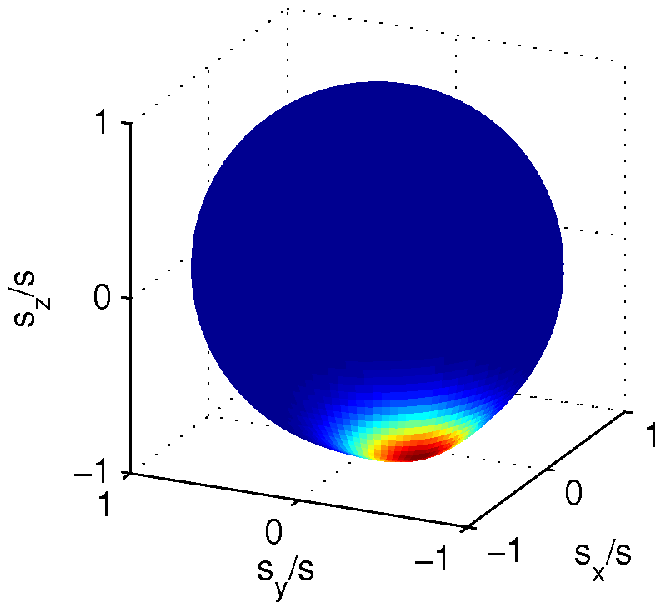}\includegraphics[scale=1]{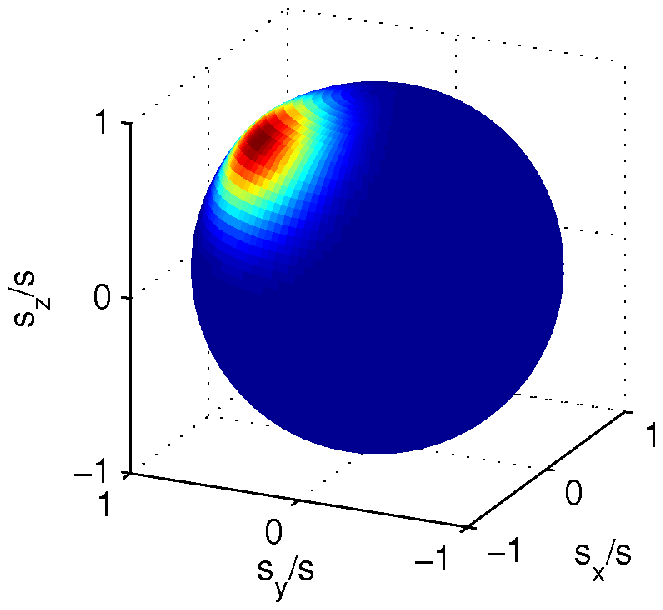}
	\caption{Husimi distributions of the island states $\left|\psi_\pm\right>$; left: $Q_-(\vartheta,\varphi) = 
	\left|\left<\vartheta,\varphi\big|\psi_-\right>\right|^2$, right: $Q_+(\vartheta,\varphi) = 
	\left|\left<\vartheta,\varphi\big|\psi_+\right>\right|^2$ for $N=20$ particles, $\varepsilon = 0$, $v=1$, $\tau=1$ 
	and $c=2/(N+1)$.}
	\label{Figure:husimi-insel}
\end{figure}

To give an impression of the localization of the Floquet states, one can compute quantum phase space distributions,
e.g.~the Husimi distributions \cite{Husi40, Hine05}
\begin{equation}
Q(\vartheta,\varphi) = \left|\left<\vartheta,\varphi\big|\psi\right>\right|^2.
\end{equation}
via the $SU(2)$ coherent states $\left|\vartheta,\varphi\right>$. In figure \ref{Figure:husimi-insel}, the Husimi distribution for the linear combinations $\left|\psi_\pm\right>$ of the two Floquet tunneling states are shown.
The sharp localization in one of the two regular islands is evident here.

\subsection{Parameter dependencies of the tunneling time}\label{subsecparameter}

In general the dynamics of the island states, however, is more complicated and shows special features. Tunneling can in 
fact be extremely stimulated or fully suppressed as we will see in the following.

\begin{figure}
	\centering
	\includegraphics[scale=1]{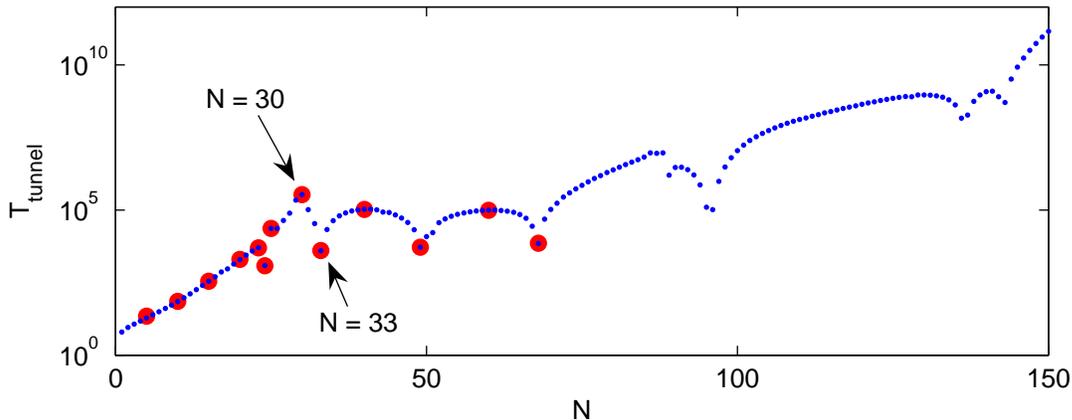}
	\caption{Tunneling period $T_{\rm tunnel}$ calculated from $\Delta\epsilon$ according to \eqref{k4_tunnelzeit} 
	as a function of $N$ for $\varepsilon = 0$, $v = 1$, $\tau=1$ and $c = 2/(N+1)$; $\textcolor{red}{\bullet}$ 
	$T_{\rm tunnel}$ determined from numerical propagation.}
	\label{Figure:tunnelzeit_c_2_N_0_150}
\end{figure}

At first we examine the tunneling for different numbers of particles. In figure \ref{Figure:tunnelzeit_c_2_N_0_150} the 
tunneling period $T_{\rm tunnel}$ for $c=2/(N+1)$ calculated with the help of \eqref{k4_tunnelzeit} is plotted over $N$;
please note the logarithmic scaling of the $T$-axis. This is in excellent agreement with the tunneling time directly 
determined from the time propagation, plotted in red for comparison.
For small $N$ the tunneling period increases exponentially. For larger $N$ new
structures appear, where the tunneling time decreases rapidly about several
orders of magnitude. These ``resonances'' have been observed for other systems
e.g.~the anharmonic driven oscillator \cite{Lin90, Shin94, Bonc98}. Quite often this phenomenon is called ``chaos assisted 
tunneling'' \cite{Lin90, Lin92, Plat92, Uter94, Haen94, Grif98} 
or ``resonance assisted tunneling'' \cite{brod01,Elts05} and can be explained as an avoided crossing of the tunneling 
doublet $\epsilon_\pm$ with some third level $\epsilon_c$ which increases the splitting $\Delta \epsilon$. In such 
regions one actually has to deal with a three state tunneling mechanism. Nevertheless the tunneling period can still 
be calculated via \eqref{k4_tunnelzeit} quite accurately.

\begin{figure}
	\centering
	\includegraphics[scale=1]{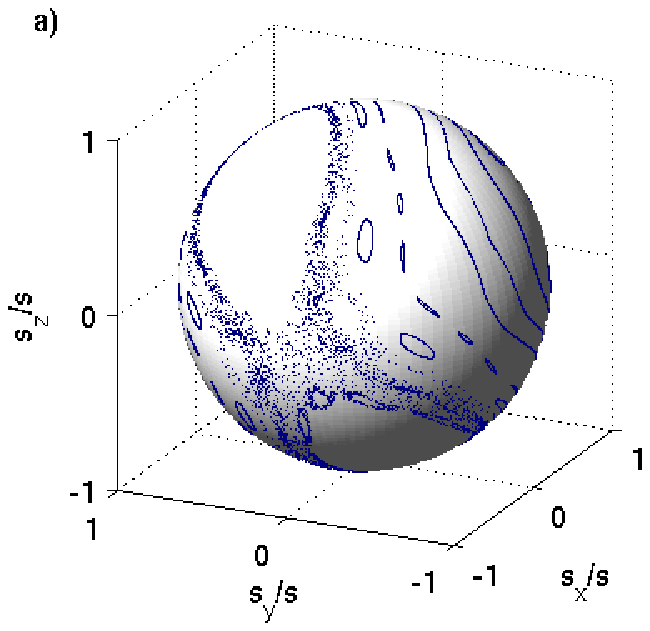}
	\includegraphics[scale=1]{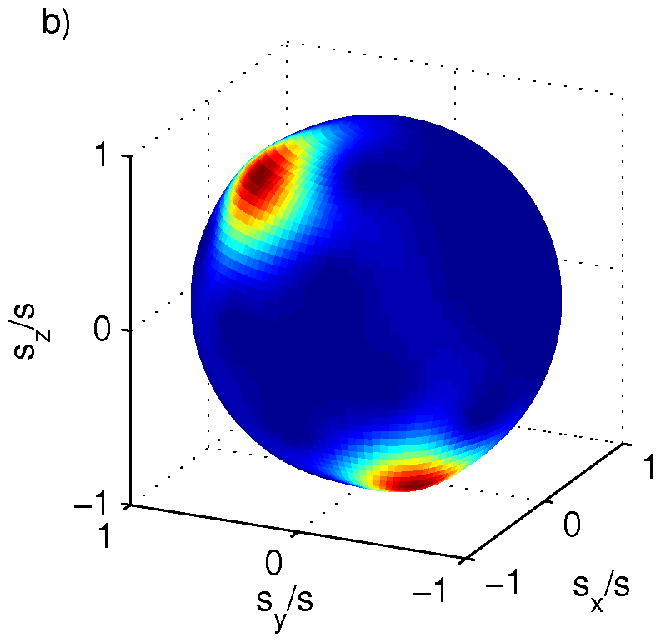}
	\includegraphics[scale=1]{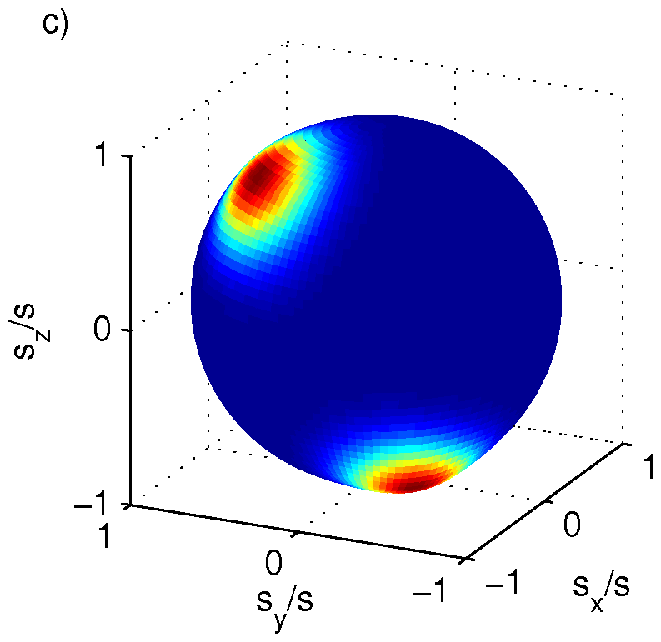}
	\includegraphics[scale=1]{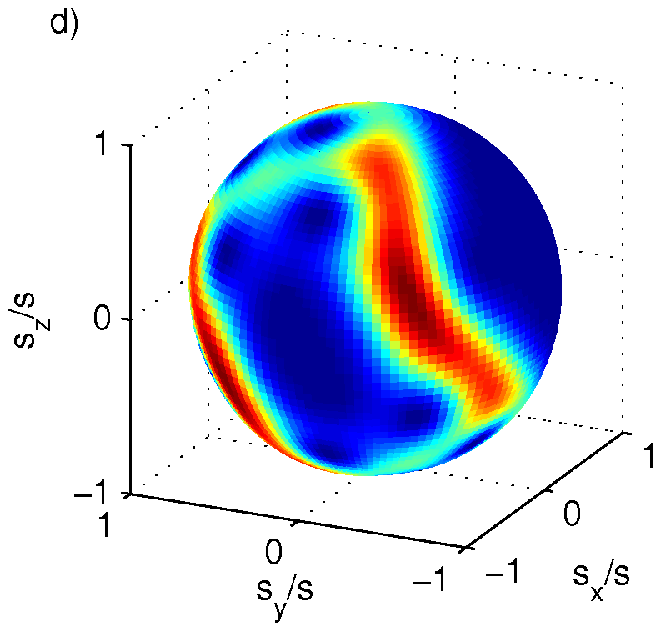}
	\caption{The three tunneling states for $N=33$, $\varepsilon = 0$, $v=1$, $\tau=1$ and $c=2/(N+1)$: a) mean-field 
	phase space , b) Husimi distribution $\left|\left<\vartheta,\varphi\big|\kappa_+\right>\right|^2$, c) 
	$\left|\left<\vartheta,\varphi\big|\kappa_-\right>\right|^2$,
	d) $\left|\left<\vartheta,\varphi\big|\kappa_c\right>\right|^2$.}
	\label{Figure:hus_N_33_c_2}
\end{figure}

\begin{figure}
	\centering
	\includegraphics[scale=1]{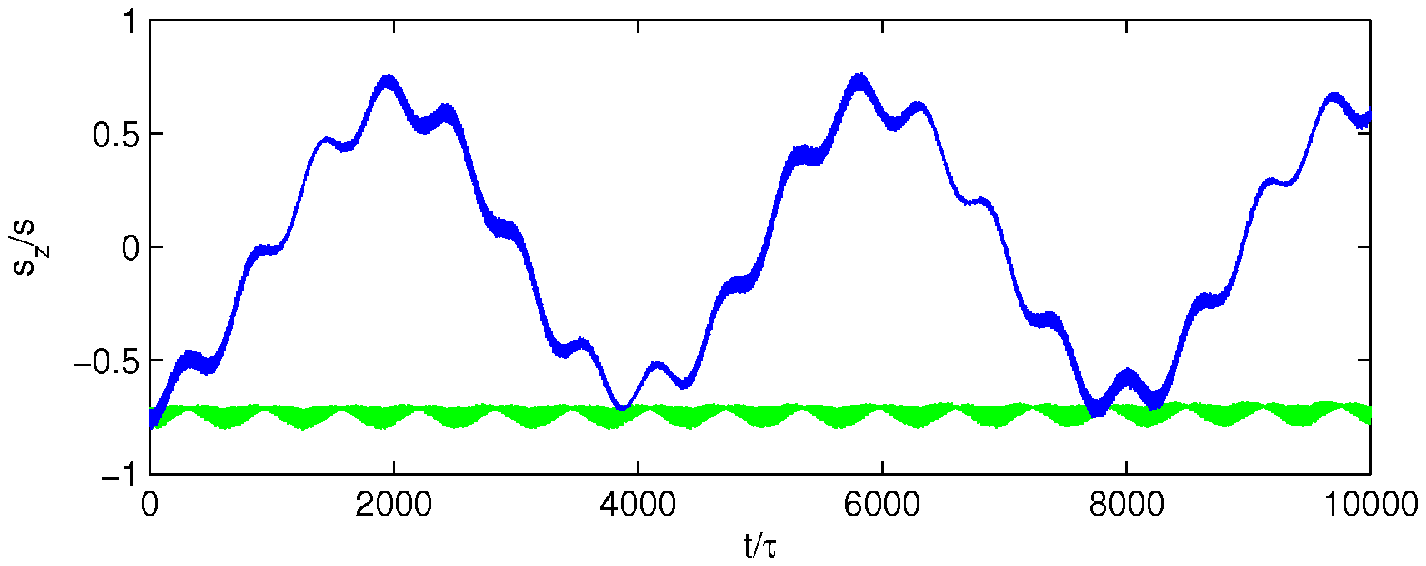}
	\caption{Population imbalance $\left<L_z\right>/\ell$ over $10000$ kick periods for $\varepsilon =0$, $v=1$
	and $c = 2/(N+1)$; $\textcolor{green}{-}$ $N=30$ particles, $\textcolor{blue}{-}$ $N=33$ particles.}
	\label{Figure:sz_N_30u33_c_2_t_10000}
\end{figure}

We illustrate this by means of the resonance at $N\approx33$ where $T_{\rm tunnel}$ changes by two orders of magnitude.
The Husimi distributions of the three Floquet states $\left|\kappa_\pm\right>$ and $\left|\kappa_c\right>$ involved in 
the tunneling for $N = 33$ are shown in figure \ref{Figure:hus_N_33_c_2}.
The third state $\left|\kappa_c\right>$ is mostly localized in the chaotic regions of the classical phase space and may 
therefore be denoted as ``chaotic'' state. The overlap of this state with the island states is actually small, 
$\left|\left<\kappa_c\big|\pm\right>\right|^2 \approx 0.06$. Nevertheless it changes the behavior of the system 
dramatically. To visualize the tunneling process, it is sufficient to focus on the $z$-component during propagation. 
In figure \ref{Figure:sz_N_30u33_c_2_t_10000} the $z$-component is shown for $N = 30$ and $N = 33$.
The great difference in the tunneling period is obvious here. The system with $N = 30$ particles shows perfect 
self-trapping up to $10^4$ kick periods, whereas the system with just three more particles exhibits strong tunneling 
with $T_{\rm tunnel} \approx 4000\tau$. Thus, the small variation of the number of particles obviously changes the 
entire tunneling process from a two state to a faster three state mechanism. The interplay with the third 
``chaotic'' state motivates the name ``chaos assisted tunneling'' (CAT) for this decrease of the tunneling period. 
If one takes a closer look on the tunneling oscillations of this system the small modulations attract attention. 
This can easily be understood as tunneling into the third state $\left|\kappa_c\right>$ which is superimposed. 
The period $T_c$ of these modulations can be calculated by the means of \eqref{k4_tunnelzeit} with the quasi-energy 
splitting $\Delta\epsilon_c$ between the third level and the tunneling doublet. In the present case this yields 
$T_c = 551$ in perfect agreement with the numerical propagation.

To achieve a better understanding of the behavior of the quasi-energies, especially crossing or avoided crossing 
scenarios in regions where the levels really get close to each other, one needs to examine their dependence on 
continuous variables, e.g., the kick strength $c$. Of special interest is the role of the third chaotic state.
One has to keep in mind, however, that the mean-field phase space structure changes by varying $c$; especially
both of the regular islands are modified.

At first we take a look at the tunneling time $T_{\rm tunnel}$ as a
function of $c$ where we leave the number of particles $N=30$ fixed.
\begin{figure}
	\centering
	\includegraphics[scale=1]{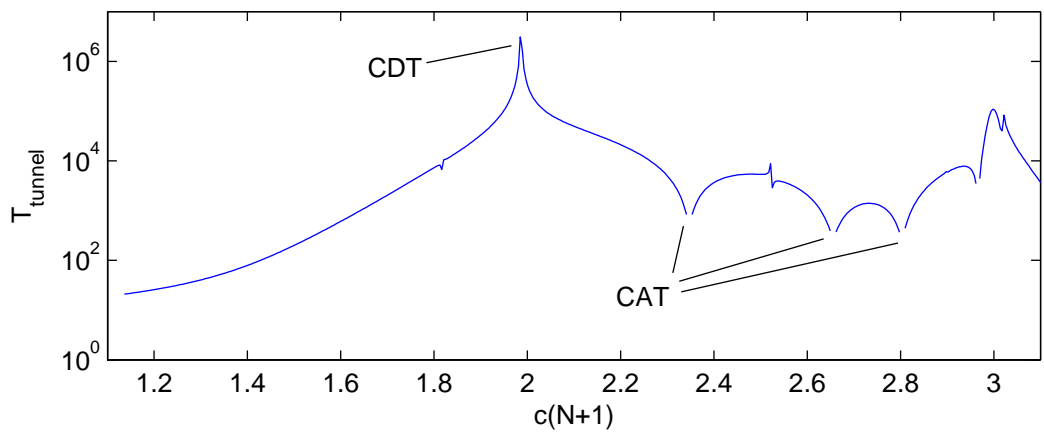}
	\caption{Tunneling time $T_{\rm tunnel}$ calculated according to \eqref{k4_tunnelzeit} with the help of
	$\Delta\epsilon$ as a function of $c$ for a fixed number of $N = 30$ particles, $\varepsilon = 0$, $v=1$
	and $\tau=1$; CDT: coherent destruction of tunneling divergences, CAT: chaos assisted tunneling resonances,
	in the regions of the small gaps the two state approximation breaks down.}
	\label{Figure:tunnel_c_1-1_3-1_N_30}
\end{figure}
This plot, shown in figure \ref{Figure:tunnel_c_1-1_3-1_N_30}, provides similar features as seen for the
$N$-dependence in figure \ref{Figure:tunnelzeit_c_2_N_0_150}. For higher values of $c$ there exist the same kind of CAT resonances as seen in the $N$ dependence above,
where the tunneling period decreases by several orders of magnitude. They are due to avoided crossing scenarios of
the quasi-energy levels, which shall be examined in more detail in the following.
\begin{figure}
	\centering
	\includegraphics[scale=1]{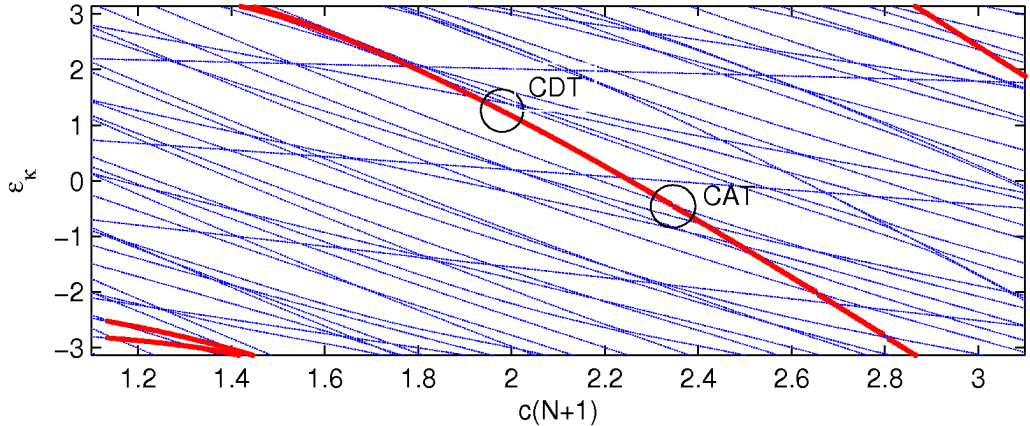}
	\caption{Quasi-energies $\epsilon_\kappa$ for $N=30$ particles, $\varepsilon = 0$, $v=1$ and $\tau=1$ as a 
	function of $c$; $\textcolor{red}{\bullet}$ tunneling doublet $\epsilon_\pm$; (CDT): real crossing of 
	$\epsilon_\pm$ inducing CDT; (CAT): avoided crossing with a chaotic level inducing CAT, see text for details.}
	\label{Figure:kreuzung00}
\end{figure}
In figure \ref{Figure:kreuzung00} we show the quasi-energies over the interaction strength $c$ for $N=30$ particles 
(note the cylindrical topology; due to the definition of the quasi-energies $-\pi$ and $\pi$ have to be identified 
on the $\epsilon$-axis). Let us focus on the prominent resonance at $c(N+1) \approx 2.35$ corresponding to the region 
marked by CAT in the figure. A magnification of this region is shown in figure \ref{Figure:kreuzung1}.
\begin{figure}
	\centering
	\includegraphics[scale=1]{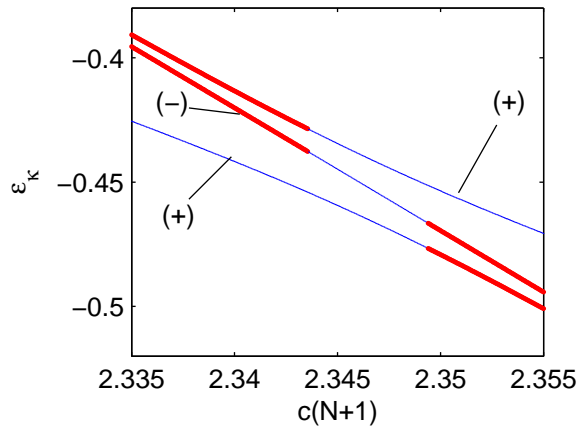}
	\caption{Blow-up of the CAT region of figure \ref{Figure:kreuzung00}; the symmetry of the quasi-energy levels
	are denoted by $(\pm)$; one observes an avoided crossing of the quasi-energy levels with $(+)$-symmetry; 
	$\textcolor{red}{\bullet}$ tunneling doublet $\epsilon_\pm$.}
	\label{Figure:kreuzung1}
\end{figure}
Here one of the levels of the doublet undergoes an avoided crossing with a chaotic third level $\epsilon_c$ (since 
they belong to the same symmetry class). The splitting of the avoided crossing is in the same order of magnitude as
the splitting of the tunneling doublet. Hence, in an intermediate region the three states $\left|\kappa_\pm\right>$
and $\left|\kappa_c\right>$ all take part in a three state tunneling mechanism. The simple two state model breaks
down here and the tunneling time cannot be evaluated according to \eqref{k4_tunnelzeit} (note the small gaps in the
graph in figure \ref{Figure:tunnel_c_1-1_3-1_N_30}). Due to the strongly avoided crossing the splitting of the
tunneling doublet increases, when the third state approaches, which causes a significant decrease of the tunneling
time. An illustration of this stimulated tunneling process, CAT, can be found in \cite{Toms94,Kohl98}.

Comparing figure \ref{Figure:tunnel_c_1-1_3-1_N_30} and \ref{Figure:kreuzung00} it becomes obvious that apart from these
resonant avoided crossing events there is a large number of avoided crossings which have no significant influence on the
tunneling period. The avoided crossing between the third level and one of the doublet levels has only a very small
splitting in these cases, which has no influence on the tunneling period, since the quasi-energy doublet stays almost
undistorted.

Despite the CAT resonance there is another structure in figure \ref{Figure:tunnel_c_1-1_3-1_N_30} which attracts attention: While for small values of $c$ the tunneling period increases faster than exponential and shows an almost monotonic behavior, a sharp peak can be observed for $c\approx 1.98/(N+1)$ which in fact is a true divergence. This phenomenon is known as ``coherent destruction of tunneling'' (CDT) \cite{Gros91a, Gros91b, Grif98}. At these points in parameter space the two levels belonging to the tunneling doublet cross. Since they have different parity such crossings are not forbidden. This zero splitting automatically leads to a divergence in the tunneling period. Thus, the relatively high value of the tunneling period that can be observed in figure \ref{Figure:tunnelzeit_c_2_N_0_150} for $N=30$ particles is due to the direct neighborhood to the CDT divergence in parameter space. A diverging tunneling time means in fact total suppression of the dynamical many particle tunneling and therefore a real self-trapping of the condensate not only in the mean-field approximation, but also in the full many particle regime.

\subsection{Tunneling landscapes}

\begin{figure}
	\centering
	\includegraphics[scale=1]{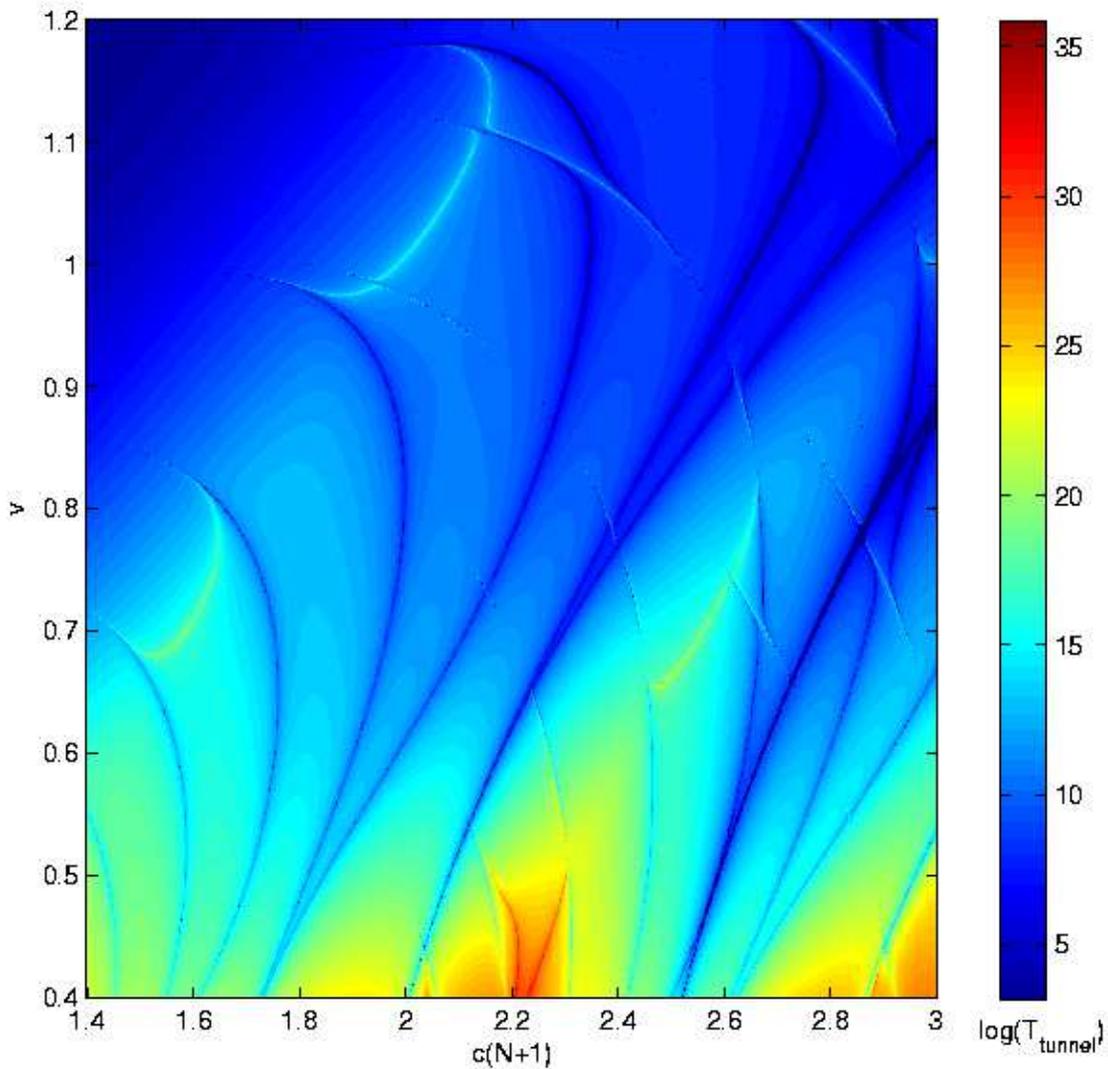}
	\caption{Tunneling landscape, $T_{\rm tunnel}$ over $(c,v)$ in false color representation on logarithmic scale; 
	$N=30$ particles, $\varepsilon = 0$ and $\tau=1$. The sharp dark lines with reduced tunneling times represent 
	CAT regions whereas these with high values of the tunneling time represent CDT regions.}
	\label{Figure:landsc0}
\end{figure}

\begin{figure}
	\centering
	\includegraphics[scale=1]{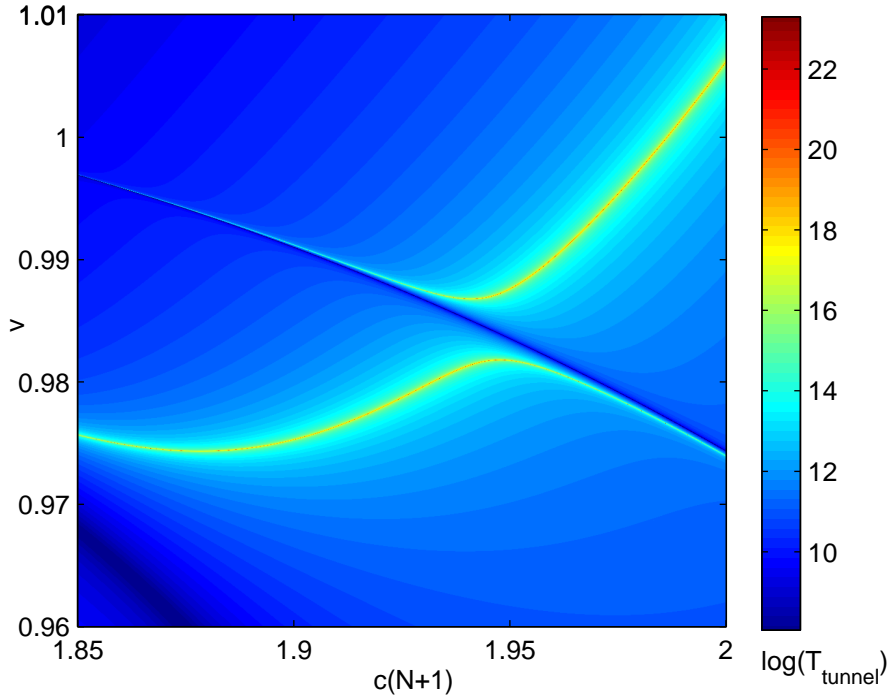}
	\caption{Tunneling landscape, blow-up of figure \ref{Figure:landsc0}; avoided crossing of two CDT ridges, 
	in the gap one observes a CAT valley.}
	\label{Figure:landsc1}
\end{figure}

Let us finally explore the whole parameter space available for a symmetric two-mode system. One system parameter is the 
interatomic
interaction strength $c$, the second parameter is the time scale of the system, which is given by the single particle
tunneling coupling $v$. Both of the effects CDT and CAT are governed by the behavior of the quasi-energies in dependence 
on the system parameters, which in turn is ruled by the symmetries of the system. These are mappings under which the 
eigenvalues and symmetry properties of $F$ are preserved; for the Floquet operator \eqref{floq} they are given by 
orthogonal matrices. Therefore $F$ has codimension $n=2$, which means that at least two system parameters have 
to be varied to enforce a degeneracy of two quasi-energies belonging to the same symmetry class \cite{Wign59,Dyso62}. 
Hence accidental degeneracies of such quasi-energy planes $\epsilon_\kappa(c,v)$  occur at most at isolated points in 
the parameter space spanned by $c$ and $v$. However, planes belonging to \textit{different} symmetry classes, 
i.e.~having opposite parity, actually exhibit exact crossings which form a one dimensional manifold in the $(c,v)$-plane.
Thus, the quasi-energy planes $\epsilon_\kappa(c,v)$ cross along lines \cite[sect. 6]{Grif98}. These lines play an 
important role in the context of tunneling since the tunneling period $T_{\rm tunnel}$ diverges there. Plotting 
$T_{\rm tunnel}$ as a function of $c$ and $v$ yields the ``tunneling landscape'' of the kicked double-well BEC. 
Figure \ref{Figure:landsc0} shows $T_{\rm tunnel}(c,v)$ in false color representation on a logarithmic scale.
One observes indeed a rich landscape with diversified sequences of ridges and valleys. The sharp lines with small 
values of the tunneling period (dark blue) correspond to the CAT regions; along these lines avoided crossings of the 
quasi-energy planes with significant splitting occur. On the other hand, the sharp lines with high values of 
$T_{\rm tunnel}$ are actually lines of divergence since the quasi-energy planes intersect along these lines in
parameter space. Thus, they correspond to the CDT regions. In the upper left corner of the presented section of
parameter space, i.e., for smaller values of $c$ and larger values of $v$, regular dynamics prevails and almost 
no crossings of the quasi-energy planes take place. The consequence of this is a flat tunneling landscape area. 
The opposite holds in the chaotic regime; here a larger number of CAT resonances and CDT divergences occur. 
Remarkable are the ``avoided crossings'' between both the ridges and the valleys of the landscape. One of these 
is shown in figure \ref{Figure:landsc1} which is a magnification of figure \ref{Figure:landsc0}.
A systematical feature at these crossings is the direct neighborhood of resonances and divergences which could 
already be observed in figure \ref{Figure:tunnel_c_1-1_3-1_N_30} that is actually a cut through figure 
\ref{Figure:landsc0} at $v=1$. Another interesting fact about this tunneling landscape is, that in principle the system 
can be tuned to the CDT regime by adjusting one parameter. Thus for this kicked system real self-trapping is possible not only in the mean-field regime, but also for the full quantum system and the system can be tuned from enforced tunneling
to this regime by rather slight variations of the parameters. This yields the possibility of a systematic and accurate population transfer between the two potential wells.

\section{Conclusions}
In the present paper we investigated a two-mode BEC with a periodically kicked interaction term in a many particle 
as well as in a mean-field description.
This system is actually equivalent to the kicked top which is a standard example of quantum chaos. Besides the 
new possible experimental realization of the kicked top, this new context sheds light on different aspects of 
the model. We showed that the tunneling oscillations of the many particle system which usually suppress the 
self-trapping effect show a rich spectrum of phenomena like coherent destruction of tunneling and chaos assisted 
tunneling. The sensitive parameter dependence of the tunneling behavior of the system offers an interesting tool 
for the manipulation of Bose-Einstein condensates.


\section*{References}
\begin{thebibliography}{10}

\bibitem{Ande95}
M.~H. Anderson, J.~R. Ensher, M.~R. Matthews, C.~E. Wieman, and W.~E. Cornell,
  Science  {\bf 269}  (1995)   198

\bibitem{Davi95}
K.~B. Davis, M.~O. Mewes, M.~R. Andrews, N.~J. van Druten, D.~S. Durfee, D.~M.
  Kurn, and W.~Ketterle,  Phys. Rev. Lett.  {\bf 75}  (1995)   3969

\bibitem{Albi05}
M.~Albiez, R.~Gati, J.~F\"olling, S.~Hunsmann, M.~Cristiani, and M.~K.
  Oberthaler,  Phys. Rev. Lett.  {\bf 95}  (2005)   010402

\bibitem{Bloc05}
I.~Bloch,  Nature Physics  {\bf 1}  (2005)   23

\bibitem{Milb97}
G.~J. Milburn, J.~Corney, E.~M. Wright, and D.~F. Walls,  Phys. Rev. A  {\bf
  55}  (1997)   4318

\bibitem{Holt01a}
M.~Holthaus and S.~Stenholm,  Eur. Phys. J. B  {\bf 20}  (2001)   451

\bibitem{Holt01b}
M.~Holthaus,  Phys. Rev. A  {\bf 64}  (2001)   011601

\bibitem{Angl01}
J.~R. Anglin and A.Vardi,  Phys. Rev. A  {\bf 64}  (2001)   013605

\bibitem{Mahm05}
K.~W. Mahmud, H.~Perry, and W.~P. Reinhardt,  Phys. Rev. A  {\bf 71}  (2005)
  023615

\bibitem{Moss06}
S.~Mossmann and C.~Jung,  Phys. Rev. A  {\bf 74}  (2006)   033601

\bibitem{Wu06}
Biao Wu and Jie Liu,  Phys. Rev. Lett.  {\bf 96}  (2006)   020405

\bibitem{06semiMP}
H.~J. Korsch and E.~M. Graefe,  Phys. Rev. A  {\bf \phantom{0}}  (2007)   in
  press (preprint: quant--ph/0611040)

\bibitem{Stoe99}
H.-J. St\"ockmann,  {\em Quantum Chaos},   Cambridge University Press,
  Cambridge, 1999

\bibitem{Haak01}
F.~Haake,  {\em Quantum Signatures of Chaos},   Springer, Berlin, Heidelberg,
  New York, 2001

\bibitem{Haak86}
F.~Haake, M.~Ku\'s, and R.~Scharf,  Z. Phys. B  {\bf 65}  (1986)   381

\bibitem{Haak00}
F.~Haake,  Journal of Modern Optics  {\bf 47}  (2000)   2883

\bibitem{Xie05}
Q.~Xie and W.~Hai,  Eur. Phys. J. D  {\bf 33}  (2005)   265

\bibitem{Zhan90}
W.-M. Zhang, D.~H. Feng, and R.~Gilmore,  Rev. Mod. Phys.  {\bf 62}  (1990)
  867

\bibitem{Arec72}
F.~T. Arecchi, E.~Courtens, R.~Gilmore, and H.~Thomas,  Phys. Rev. A  {\bf 6}
  (1972)   2211

\bibitem{Gilm75}
R.~Gilmore, C.~M. Bowden, and L.~M. Narducci,  Phys. Rev. A  {\bf 12}  (1975)
  1019

\bibitem{Pere86}
A.~M. Perelomov,  {\em Generalized Coherent States and Their Applications},
  Springer, Berlin Heidelberg New York London Paris Tokyo, 1986

\bibitem{07meanf}
D.~Witthaut, F.~Trimborn, and H.~J. Korsch,  {\bf \phantom{0}}  (2007)
  (preprint: cond--mat/0701383)

\bibitem{Grif98}
M.~Grifoni and P.~H\"anggi,  Phys. Rep.  {\bf 304}  (1998)   229

\bibitem{Bayf99}
J.~E. Bayfield,  {\em Quantum Evolution},   John Wiley and Sons, New York, 1999

\bibitem{Lin90}
W.~A. Lin and L.~E. Ballentine,  Phys. Rev. Lett.  {\bf 65}  (1990)   2927

\bibitem{Lin92}
W.~A. Lin and L.~E. Ballentine,  Phys. Rev. A  {\bf 45}  (1992)   3637

\bibitem{Bavl93}
R.~Bavli and H.~Metiu,  Phys. Rev. A  {\bf 47}  (1993)   3299

\bibitem{95tun}
V.~Averbuckh, N.~Moiseyev, B.~Mirbach, and H.~J. Korsch,  Z. Phys. D  {\bf 35}
  (1995)   247

\bibitem{Bonc98}
L.~{Bonci, A. Farusi, P. Grigolini and R. Roncaglia},  Phys. Rev. E  {\bf 58}
  (1998)   5689

\bibitem{Aver02}
V.~Averbukh, Shmuel Osovski, and N.~Moiseyev,  Phys. Rev. Lett.  {\bf 89}
  (2002)   253201

\bibitem{Hens00}
W.~K. Hensinger, A.~G. Truscott, B.~Upcroft, N.~R. Heckenberg, and
  H.~Rubinsztein-Dunlop,  J. Opt. B  {\bf 2}  (2000)   659

\bibitem{Hens01a}
W.~K.~Hensinger et. al.,  Nature  {\bf 412}  (2001)   52

\bibitem{Hens01b}
W.~K.~Hensinger et. al.,  Phys. Rev. A  {\bf 64}  (2001)   033407

\bibitem{Stec01}
D.A. Steck, W.~H. Oskay, and M.~G. Raizen,  Science  {\bf 293}  (2001)   274

\bibitem{Salm02}
G.L. Salmond, C.~A. Holmes, and G.~J. Milburn,  Phys. Rev. A  {\bf 65}  (2002)
   033623

\bibitem{Osov05}
S.~Osovski and N.~Moiseyev,  Phys. Rev. A  {\bf 72}  (2005)   033603

\bibitem{Husi40}
K.~Husimi,  Proc. Phys. Math. Soc. Japan  {\bf 22}  (1940)   264

\bibitem{Hine05}
A.~P. Hines, R.~H. McKenzie, and G.~J. Milburn,  Phys. Rev. A  {\bf 71}  (2005)
    042303

\bibitem{Shin94}
J.~Y. Shin and H.~W. Lee,  Phys. Rev. E  {\bf 50}  (1994)   902

\bibitem{Plat92}
J.~Plata and J.~M.~Gomez Llorente,  J. Phys. A  {\bf 25}  (1992)   L303

\bibitem{Uter94}
R.~Utermann, T.~Dittrich, and P.~H\"anggi,  Phys. Rev. E  {\bf 49}  (1994)
  273

\bibitem{Haen94}
P.~H\"anggi, R.~Utermann, and T.~Dittrich,  Physica B  {\bf 194-196}  (1994)
  1013

\bibitem{brod01}
O.~Brodier, P.~Schlagheck, and D.~Ullmo,  Phys. Rev. Lett.  {\bf 87}  (2001)
  064101

\bibitem{Elts05}
C.~Eltschka and P.~Schlagheck,  Phys. Rev. Lett.  {\bf 94}  (2005)   014101

\bibitem{Gros91a}
F.~Grossmann, T.~Dittrich, P.~Jung, and P.~H\"anggi,  Phys. Rev. Lett.  {\bf
  67}  (1991)   516

\bibitem{Gros91b}
F.~Grossmann, P.~Jung, T.~Dittrich, and P.~H\"anggi,  Z. Phys. B  {\bf 84}
  (1991)   315

\bibitem{Toms94}
S.~Tomsovic and D.~Ullmo,  Phys. Rev. E  {\bf 50}  (1994)   145

\bibitem{Kohl98}
S.~{Kohler, R. Utermann, P. H\"anggi and T. Dittrich},  Phys. Rev. E  {\bf 58}
  (1998)   7219

\bibitem{Wign59}
E.~P. Wigner,  {\em Group Theory and its Applications to the Quantum Mechanics
  of Atomic Spectra},   Academic, New York, 1959

\bibitem{Dyso62}
F.~J. Dyson,  J. Math. Phys.  {\bf 3}  (1962)   140, 157, 166

\end{thebibliography}

\end{document}